\documentclass[aps,prl,twocolumn,floats,showpacs]{revtex4}
  \usepackage{epsfig}
  \usepackage{amsmath}
  \usepackage{amssymb}
\begin{document}
  \title{Two-channel Kondo tunneling in triple quantum dot}
  \author{
  T. Kuzmenko$^1$, K. Kikoin$^1$ and Y. Avishai$^{1,2}$\\
  }
  \affiliation {$^1$Department of Physics and $^2$Ilse Katz Center,
  Ben-Gurion University, Beer-Sheva, Israel }
  \date{\today}
\begin{abstract}
The effective spin Hamiltonian of
a triple quantum dot with odd electron occupation 
weakly connected in series with
left ($l$) and right ($r$) metal leads is composed of 
two-channel exchange and co-tunneling terms. 
Renormalization group equations for the corresponding three 
exchange constants $J_{l}$, $J_{r}$ and $J_{lr}$ are
solved (to third order).
Since $J_{lr}$ is relevant, the system is 
mapped on an anisotropic two-channel Kondo problem. 
The structure of the conductance as function of
temperature and gate voltage implies that 
in the weak and intermediate coupling regimes, 
two-channel Kondo physics  
persists at temperatures as low as several $T_{K}$.
At even electron occupation, the number of channels equals
twice the spin of the triple dot (hence it is a fully screened impurity).
\end{abstract}
\maketitle

\noindent {\bf Motivation}: In the present work, a simple
configuration of localized moment in nanostructures is studied,
where the two-channel Kondo Hamiltonian appears in resonance
tunneling. Concrete experiment is proposed in order to elucidate
the pertinent physics at $T>T_{K}$ (the Kondo temperature). In the
strong coupling regime, a multichannel Kondo system is known to be
a non-Fermi liquid \cite{nobl80}, but construction of simple
theoretical models pertaining to experimentally feasible setups
 is notoriously elusive. Examples are
magnetic impurity scattering, physics of two-level systems and
Kondo lattices (see \cite{schlot93,coza99} for review). Recent
attempts to realize two-channel Kondo effect in tunneling through
quantum dots \cite{Oreg} using peculiar setups still await
experimental manifestation. The problem is exemplified in
tunneling through a simple quantum dot sandwiched between two
metallic "left" ($l$) and "right" ($r$) leads. Starting from the
single impurity Anderson model, it is tempting to think of the two
leads as a source of two tunneling channels. However, if the two
fermion lead operators $c_{k \sigma a}$ ($k=$ momentum, $\sigma=$
spin projection and $a=l,r$ the lead index) are coupled to a
single dot electron operator $d_{\sigma}$, one channel can always
be eliminated by an appropriate rotation in $l-r$ space
\cite{Glaz}:
\begin{eqnarray}
&& c_{k \sigma l}=\cos\theta_k c_{k \sigma 1}
+\sin \theta_k c_{k \sigma 2}, \nonumber \\
&& c_{k \sigma r}=-\sin\theta_k c_{k \sigma 1} +\cos \theta_k c_{k
\sigma 2}. \label{GR}
\end{eqnarray}
As a result, only the standing wave fermions $c_{k \sigma 1}$
contribute to tunneling. It is therefore natural to expect that a
generic two-channel Hamiltonian can be realized only when the
rotation (\ref{GR}) cannot eliminate the second channel. Such
situation may arise, e.g., when, instead of a simple quantum dot
one has a nanoobject consisting of several dots so that $c_{k
\sigma l}$ and $c_{k \sigma r}$
 are coupled to {\it different} dot
electron operators.

\noindent {\bf Model Hamiltonian}: Consider a triple quantum dot
(TQD) which consists of three dots $l,f,r,$ connected in series to
left and right leads (see Fig. 1). The figure defines also
tunneling amplitudes $V_{a}$ and hopping amplitudes $W_{a}$
between the side dots and the central one. The source of Kondo
screening is a Coulomb blockade in the central dot $f$, which is
supposed to have a smaller radius (and hence a larger capacitive
energy) than the two side dots, namely, $Q_f\gg Q_{l,r}$ (cf.
\cite{KKA}). The tunneling Hamiltonian has the form $
H_t=\sum_{a=l,r}\sum_{k \sigma}V_{a}c^\dagger_{k \sigma
a}d_{\sigma a}+H.c., $ and the rotation (\ref{GR}) does not
eliminate the second tunneling channel.
 Note that direct tunneling
through the TQD is suppressed due to electron level mismatch and
Coulomb blockade, so that {\it only co-tunneling mechanism}
contributes to the current.
\begin{figure}[htb]
\centering
\includegraphics[width=70mm,height=70mm,angle=0]{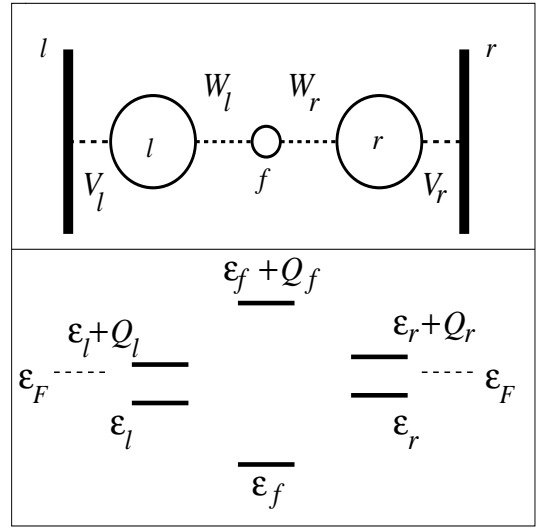}
\caption{Triple quantum dot in series (upper panel) 
and single-electron energy levels of each individual 
dot $\varepsilon_{a}=\epsilon_{a}-V_{ga}$=bare energy
minus gate voltage (lower panel).}
\label{TQD}
\end{figure}
To quantify these elementary statements consider the case of TQD
with three electrons. The Hamiltonian of the isolated TQD reads
\begin{eqnarray}
&& H_{d}=\sum_{\alpha=l,r,f}\left(\sum_{\sigma}
\varepsilon_{\alpha} n_{\sigma \alpha} + Q_{\alpha} n_{\uparrow
\alpha}n_{\downarrow \alpha}
\right) \nonumber \\
&& + \sum_{a=l,r}\left( W_a d^\dagger_{\sigma a}d_{f\sigma} +
H.c.\right), \label{Hdot}
\end{eqnarray}
where $n_{\sigma a}=d^\dagger_{\sigma a}d_{\sigma a}$, with
$\sum_{\sigma a} n_{\sigma a}=3.$ This Hamiltonian can be easily
diagonalized in the space of three-electron states $|\Lambda
\rangle$ of the TQD. The lowest ones are classified as a ground
state doublet $|d_1\rangle$, low-lying doublet excitation
$|d_2\rangle$ and quartet excitation $|q\rangle$. The {\it single
electron} levels $\varepsilon_a=\epsilon_{a}-V_{ga}$ are tuned by
gate voltages $V_{ga}$ such that $\beta_{a} \equiv
W_{a}/\Delta_{a}\ll 1$ $(\Delta_{a} \equiv
\varepsilon_{a}-\varepsilon_f)$. To order $\beta_a^2$, the
corresponding energy levels $E_\Lambda$ are
\begin{eqnarray}
&& E_{d_1}={\varepsilon}_f +{\varepsilon}_l +{\varepsilon}_r-
\frac {3} {2} \left[\frac{W_l^2}{\Delta_l}+
\frac{W_r^2}{\Delta_r}\right], \nonumber \\
&&E_{d_2}=E_{d_1}+ \left[\frac{W_l^2}{\Delta_l}+
\frac{W_r^2}{\Delta_r}\right], \nonumber \\
&&E_{q}={\varepsilon}_f +{\varepsilon}_l +\varepsilon_r.
\label{Edot}
\end{eqnarray}
The states $|d_{1}\rangle$ and $|d_{2} \rangle$ include components
with two electrons on $l$ or $r$ dots (responsible for
co-tunneling). Coupling with the leads admixes
 three-particle states $|\Lambda \rangle$ of the dot,
with two and four particle states $|\lambda \rangle$ (for brevity
we include only the formers, addition of the latters is
straightforward). It is then useful to introduce diagonal and
number changing dot Hubbard operators
$X^{\Lambda\Lambda}=|\Lambda\rangle\langle\Lambda|$ and
$X^{\lambda\Lambda}=|\lambda\rangle\langle\Lambda|$ respectively.
The Hamiltonian of the whole system (leads, TQD and tunneling)
then reads
\begin{eqnarray}
&& H=\sum_{ k \sigma a}\epsilon_{ka} c^{\dagger}_{k \sigma a}c_{k
\sigma a} + \sum_\Lambda {E}_\Lambda X^{\Lambda\Lambda}
+ \sum_\lambda {E}_\lambda X^{\lambda\lambda}  \nonumber \\
&& + \left(\sum_{\Lambda\lambda}\sum_{k \sigma
a}V^{\Lambda\lambda}_{\sigma a} c^{\dagger}_{k \sigma
a}X^{\lambda\Lambda}+ H.c.\right). \label{H0}
\end{eqnarray}
Here $-\frac {D_a} {2}< \epsilon_{ka}< \frac {D_a} {2}$ are lead
electron energies with nearly identical bandwidths $D_l\approx D_r
\equiv D_1$. Within linear response, the Fermi energies are the
same on both sides, $\epsilon_{Fl}=\epsilon_{Fr}=\epsilon_{F}=0$.
The tunneling amplitudes
 $V^{\Lambda\lambda}_{\sigma a}\equiv
V_a\langle\lambda|d_{\sigma a}|\Lambda\rangle$ depend explicitly
on the lead index $a$ and on the respective 2-3 particle quantum
numbers $\lambda,\Lambda$. Elimination of one of the two channels
by the rotation (\ref{GR}) is then impossible: adding $l$ or $r$
electron to a given state $|\lambda\rangle$ results in different
states $|\Lambda\rangle$.

\noindent {\bf RG procedure and spin Hamiltonian}: Following a
two-stage RG procedure the low-energy Kondo tunneling through TQD
may be exposed. First, the bandwidth is continuously reduced from
$D_1$ to $D_{0}<D_1$ (see \cite{KKA} and references therein), and
the energy levels (\ref{Edot}) are renormalized by eliminating
high-energy charge excitations \cite{Hald}. If at the end of this
procedure $E_{d_1}(D_{0}) < -D_{0}/2$ and
$E_{d_2}(D_{0})-E_{d_1}(D_{0})$ exceeds $T_K$ (defined below),
charge fluctuations are quenched and the doublet ground state
$|d_{1\sigma=\uparrow,\downarrow} \rangle$ becomes a localized
moment. Following the Schrieffer-Wolff transformation the spin
Hamiltonian reads
\begin{eqnarray}
&&H_{s} = J_l{\bf S}\cdot {\bf s}_l+J_r{\bf S}\cdot {{\bf
s}_r}+J_{lr} {\bf S}\cdot({\bf s}_{lr}+{\bf s}_{rl}). \label{Hs}
\end{eqnarray}
The spin 1/2 operator ${\bf S}$ acts on
$|d_{1\sigma=\uparrow,\downarrow} \rangle$ whereas the lead
electrons spin operators are ${\bf s}_{a}= \sum_{kk'} c_{k \mu
a}^{\dagger} {\boldsymbol \sigma}_{\mu \nu} c_{k' \nu a}$. The
presence of the left-right spin operator ${\bf s}_{lr}= \sum_{kk'}
c_{k \mu l}^{\dagger} {\boldsymbol \sigma}_{\mu \nu} c_{k' \nu r}$
is responsible for co-tunneling current. Moreover, $J_{a}=8
{\gamma}^2 |V_{a}|^2/3 (\epsilon_F-\varepsilon_{a})>0$ (where
$\gamma=\sqrt {1-\frac {3} {2} (\beta_{l}^{2}+\beta_{r}^{2})}$),
while $J_{lr}=-4 V_{l} V_{r} \beta_{l} \beta_{r}/ 3(\epsilon_{F}-
\varepsilon_{f})$ so that $|J_{lr}/J_{a}|$ is of order $\beta_{l}
\beta_{r} \ll 1$. The Hamiltonian (\ref{Hs}) then encodes a
two-channel Kondo physics, where the leads serve as two
independent channels and $T_{K}=max\{T_{Kl},T_{Kr}\}$,
$T_{Ka}=D_{0}e^{-D_{0}/J_{a}}$.

\begin{figure}[htb]
\centering
\includegraphics[width=80mm,height=50mm,angle=0,]{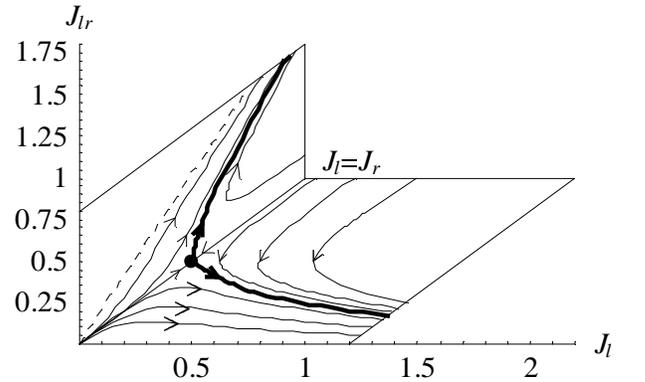}
\caption{Scaling trajectories for two-channel Kondo effect in
TQD.} \label{flow}
\end{figure}

In the second stage of the RG procedure, a poor-man scaling
technique is used to renormalize the exchange constants by further
reducing the band-width $D_{0} \to D$. The pertinent fixed points
are then identified as $D \to T_{K}$ \cite{Anderson}. Unlike the
situation encountered in the single-channel Kondo effect, third
order diagrams in addition to the usual single-loop ones should be
included (see Fig.5 in Ref. \cite{nobl80} and Fig.9 in Ref.
\cite{Coleman}).
Below we use $D_{0}$ as an energy unit, hence $J_a$, $V_a$, $W_a$,
$V_{ga}$, $\varepsilon_a$, $T$, $D$ {\it etc.} now become
dimensionless. With $a=l,r$ and ${\bar a}=r,l$ the three RG
equations for $J_l, J_r, J_{lr}$ are,
\begin{eqnarray}
&& \frac {d J_{a}} {d \ln D}=-(J_{a}^{2}+J_{lr}^2)+
J_{a}(J_{a}^{2}+J_{\bar{a}}^{2}+ 2 J_{lr}^{2}), \nonumber \\
&&\frac{d J_{lr}}{d\ln D}= -J_{lr}\left(J_l+J_r
\right)+J_{lr}(J_l^2+J_r^2+ 2J_{lr}^2). \label{RG3}
\end{eqnarray}
On the symmetry plane $J_l=J_r \equiv J$, equations (\ref{RG3})
reduce to a couple of RG equations for $J_{1,2}=J \pm J_{lr}$
\begin{equation}
dJ_i/d\ln D= -J_i^2+ J_i(J_1^2 + J_2^2) \ \ \ (i=1,2), \label{RG4}
\end{equation}
subject to $J_{i}(D_{0}=1)=J_{i0}$. These are the well-known
equations for the anisotropic two-channel Kondo
effect\cite{nobl80}. With $\phi_{i} \equiv (J_{1}+J_{2}-1)/J_{i}$,
$C_i \equiv \phi_{{\bar i} 0}-\phi_{i0}$ and $L_i(x) \equiv
x-\ln(1+C_i/x)-2 \ln x$, the solution of the system (\ref{RG4}) is
\begin{eqnarray}
&& L_i(\phi_{i})-L_i(\phi_{i0})=-\ln D \ \ \ (i=1,2). \label{RGS}
\end{eqnarray}
The scaling trajectories in the sector $(J_l\geq J_r \geq
0,J_{lr}=0)$ and in the symmetry plane with $0<J_{lr}<J$ are shown
in Fig.~\ref{flow}.
Although the fixed point (1/2, 1/2, 0) remains inaccessible if
$J_{lr} \ne 0$, one may approach it close enough starting from an
initial condition $J_{lr0}\ll J_{l0},J_{r0}$. Realization of this
inequality is a generic property of TQD in series shown in Fig.1.\\
{\bf Conductance}: According to general perturbative expression
for the dot conductance \cite{kang00}, its zero-bias anomaly is
encoded in the third order term,
\begin{eqnarray}
&& G^{(3)}=G_0 J_{lr}^{2}\left[J_{l}(T) + J_{r}(T) \right], \ \ \
(G_{0}=\frac{2 e^2} {h}). \label{Gpeak}
\end{eqnarray}
Here the temperature $T$ replaces the bandwidth $D$ in the
solution (\ref{RGS}). Let us present a qualitative discussion of
the conductance $G[J_{a}(T)]$ (or in an experimentalist friendly
form, $G(V_{ga},T)$) based on the flow diagram \ref{flow}.
(Strictly speaking, the RG method and hence the discussion below,
is mostly reliable in the weak coupling regime $T > T_{K}$).
Varying $T$ implies moving on a curve
$[J_{l}(T),J_{r}(T),J_{lr}(T)]$ in three dimensional parameter
space (Fig. \ref{flow}), and the corresponding values of the
exchange parameters determine the conductance according to
equation (\ref{Gpeak}). Note that if, initially,
$J_{l0}=J_{r0}\equiv J_{0}$ the point will remain on a curve
$[J(T),J(T),J_{lr}(T)]$ located on the symmetry plane. By varying
$V_{ga}$ it is possible to tune the initial condition
$(J_{l0},J_{r0})$ from the highly asymmetric case $J_{l0} \gg
J_{r0}$ to the fully symmetric case $J_{l0}=J_{r0}$. For a fixed
value of $J_{lr0}$ the conductance shoots up (logarithmically) at
a certain temperature $T^{*}$ which decreases toward $T_{K}$ with
$|J_{l0}-J_{r0}|$ and $J_{lr0}$. The closer $T^{*}$ is to $T_{K}$,
the closer is the behavior of the conductance to that expected in
a generic two-channel situation. Thus, although the isotropic
two-channel Kondo physics is unachievable in the strong coupling
limit, its precursor might show up in the 
intermediate coupling regime.

\begin{figure}[htb]
\centering
\includegraphics[width=60mm,height=45mm,angle=0,]{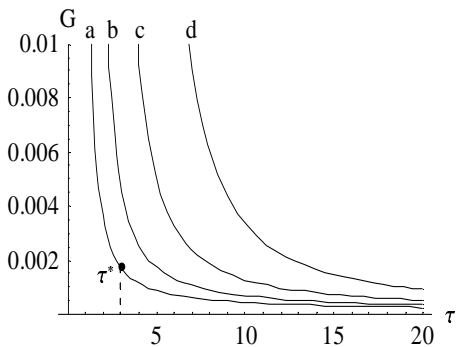}
\caption{Conductance $G$ in units of $G_{0}$ as a function of
temperature ($\tau=T/T_{K}$), at various gate voltages. The lines
correspond to:
 (a) the symmetric case $J_l=J_r$ ($V_{gl}=V_{gr}$),
(b-d) $J_l \gg J_r$, with $V_{gl}-V_{gr}=0.03,\ 0.06$ and $0.09$.
At $\tau \to \infty$ all lines converge to the bare conductance.}
\label{GT}
\end{figure}

\begin{figure}[htb]
\centering
\includegraphics[width=60mm,height=50mm,angle=0,]{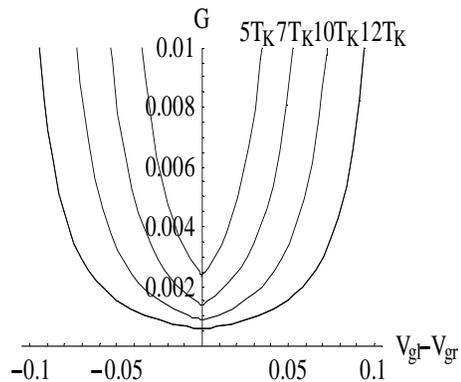}
\caption{Conductance $G$ in units of $G_{0}$ as a function of gate
voltage at various temperatures (at the origin $J_{l}=J_{r}$).}
\label{GV}
\end{figure}

The conductance $G(V_{ga},T)$ as function of $T$ for several
values of $V_{ga}$ and the same value of $J_{lr0}$ is displayed by
the family of curves in Fig.\ref{GT}. For $G$ displayed in
curve $a$, $T^{*}/T_{K}\approx 3$ and for $T > T^{*}$ it is very
similar to what is expected in an isotropic two-channel system.
Alternatively, holding $T$ and changing gate voltages $V_{ga}$
enables an experimentalist to virtually cross the symmetry plane.
This is equivalent to moving vertically downward on Fig.
\ref{GT}. At high temperature the curves almost coalesce and the
conductance is virtually flat. At low temperature (still above
$T_{K}$) the conductance exhibits a sharp minimum. This is
summarized in Fig.\ref{GV}.

\noindent {\bf Higher degeneracy and dynamical symmetries}: The
spin Hamiltonian (\ref{Hs}) is expressible in terms of the dot
(ground-state) spin ${\bf S}=1/2$  components
 $S_{i}$ ($i=x,y,z$), the generators of the group
$SU(2)$. Due to complete spin degeneracy $E_{d_{1}
\uparrow}=E_{d_{1} \downarrow}$, the dot Hamiltonian itself is of
course invariant under $SU(2)$ but the hybridization leading to
the exchange terms $J_{a} {\bf S} \cdot {\bf s}_a$ clearly breaks
it, allowing for dot spin-flips. In composite quantum dots, the
degeneracy of spin states at the end of the first stage RG
procedure might be much richer \cite{KKA} (and greater than 2).
Our analysis shows that for the present model of TQD with $N=3$
electrons there exists a scenario of level degeneracy in which the
renormalized energies of the two doublets and the quartet are
degenerate at the Schrieffer-Wolff limit, that is,
$E_{d_1}(D_{0})=E_{d_2}(D_{0})=E_{q}(D_{0})\le -D_{0}/2$. The
corresponding wave functions $|\Lambda\rangle$ are vector sums of
states composed
 of a "passive" electron sitting in the central dot
and singlet/triplet (S/T) two-electron states in the $l,r$ dots.
Using certain combinations of dot Hubbard operators $|\Lambda
\rangle \langle \Lambda' |$ one can now define two vector
operators ${\bf S}$ and ${\bf M}$ such that ${\bf S}$ is the dot
spin 1 operator responsible for transitions within the triplet
states, while  ${\bf M}$ accounts for S/T transitions\cite{KKA}.
The operators ${\bf S}$ and ${\bf M}$ together with the spin $1/2$
operator ${\bf s}$ of the central dot electron generate the group
$SO(4) \times SU(2)$ specified by the Casimir operator
$S^{2}+M^{2}+s^{2}=\frac {15} {4}$. The corresponding spin
Hamiltonian,
\begin{equation}
H_{s}=\sum_{a=l,r} [ J_{a}^{T}{\bf S} +J_{a}^{ST} {\bf M}
 ]\cdot {\bf s}_{a} +J_{lr} {\bf s}\cdot({\bf s}_{lr}+{\bf
s}_{rl}), \label{HSO4}
\end{equation}
expressible in terms of its generators breaks that symmetry. Here
$J_{a}^{T}=4 \gamma
|V_{a}|^{2}/3(\epsilon_{F}-\varepsilon_{a})>0$,
$J_{a}^{ST}=\sqrt{1-\frac{1}{2}\left(\beta_l^2+\beta_r^2\right)}
J_{a}^{T}$ and $J_{lr}=-4 V_{l} V_{r} \beta_{l} \beta_{r}
/(\epsilon_{F}-\varepsilon_{f})$, so that $|J_{lr}|/J_{a}^{T}
\approx \beta^{2}$. Both ${\bf S}$ and ${\bf M}$ vectors are
involved in Kondo screening, a situation much richer than the one
described by the single impurity Hamiltonian (\ref{Hs}).

\noindent As far as the Kondo physics is concerned, the number of
channels $n=2$ exactly equals twice the spin, $2S$. Therefore, the
physics in the strong coupling limit is similar to that of the
{\it single} channel one. The two stable fixed points are
$(J_{a}^{T},J_{a}^{ST},J_{\bar a}^{T},J_{\bar a}^{ST})=
(\infty,\infty,0,0)$ implying $T_{K}=max \{T_{Kl},T_{Kr}\},$
$T_{Ka}=\left\{ \exp{[-1/(J_{a}^{T}+J_{a}^{ST})]} \right\}.$ In
the weak coupling limit, systems with $2S>n$ and $2S=n$ might have
different physics.

Finally, let us discuss the experimentally relevant situation of
changing electron occupation $N$ in the same TQD device. For
$N=2$, the lowest--energy states consist of two singlets $|S_{a}
\rangle$ and two triplets $|T_{a} \rangle$ (one electron in the
left and right dot and one electron in the central dot). The next
charged transfer excitons consist of a singlet $|S_{lr}\rangle $
and a triplet $|T_{lr} \rangle$ (one electron in the left and
right dots). The corresponding energy levels are
\begin{eqnarray}
E_{S_{a}} &=& {\varepsilon}_f +{\varepsilon}_{a}
 -\frac{2W_{a}^2}{\Delta_a+Q_a}-W_{\bar a}\beta_{\bar a},
\nonumber\\
E_{{T_a}} &=& {\varepsilon}_f +{\varepsilon}_{a}-W_{\bar
a}\beta_{\bar a},
\nonumber\\
E_{S_{lr}} &=& E_{T_{lr}} = {\varepsilon}_{l} +{\varepsilon}_{r}
+W_{l}\beta_{l}+ W_{r}\beta_{r}. \label{En}
\end{eqnarray}
where $\beta_a=W_a/\Delta_a\ll 1$,
$(\Delta_a=\varepsilon_a-\varepsilon_f)$.

The most symmetric case is realized when
${\varepsilon_l}={\varepsilon_r} \equiv \varepsilon$ and
${V_{l}}={V_{r}} \equiv V.$ If at the end of the first stage RG
procedure one has $E_{S_{l}}\simeq E_{T_{l}}\simeq E_{S_{r}}\simeq
E_{T_{r}}$, the TQD possesses a $P\times SO(4)\times SO(4)$
symmetry (with $P$ is $l\leftrightarrow r$ permutation). Following
an RG procedure and a Schrieffer-Wolff transformation, the spin
Hamiltonian reads,
\begin{eqnarray}
{H} &=& J^T_1 \sum_{a} {\bf S}_{a}\cdot {{\bf s}_a}+
J^{ST}_1 \sum_{a} {\bf M}_{a}\cdot {{\bf s}_a}\nonumber\\
&+& {J_{2}^{T}}{\hat P}\sum_{a}{\bf S}_a\cdot {\bf s}_{a{\bar a}}+
J_{2}^{ST}{\hat P}\sum_{a}{\bf M}_a\cdot {\bf s}_{a{\bar
a}}\nonumber\\
&+&\sum_{a=l,r}\left[J^T_{lr}{\bf S}_{a}+J^{ST}_{lr}{\bf
M}_{a}\right]\cdot ({\bf s}_{lr}+{\bf s}_{rl}), \label{Hren}
\end{eqnarray}
where $J^T_{1}(D_0)=J^T_2(D_0)= \frac{2 (1-
\beta^{2})|V|^2}{\epsilon_{F}-\varepsilon},$
$J^T_{lr}(D_0)=\frac{2
\beta^{2}|V|^2}{\epsilon_{F}-\varepsilon_f},$ and
$J^{ST}(D_0)=\sqrt{1-\frac{2 W^{2}}{(\Delta+Q)^{2}}} J^T(D_0)$,
and ${\hat P}=\sum_{a=l,r}(X^{T_aT_{\bar a}}+X^{S_aS_{\bar a}})$.
 Here again, the Kondo physics falls
under the category $2S=n$. It is then sufficient to write (and
solve) second order poor-man scaling equations. Having done it, we
find the stable fixed points $J_{i}^{T},J_{i}^{ST} \to \infty$
($i=1,2$). These define the corresponding Kondo temperature
$$T_{K}=\exp\left(-\frac{2}
{(\sqrt{3}+1)(J_1^{T}+J_1^{ST})}\right).$$
For $N=4$, similar analysis shows that the TQD again manifests the
fully screened Kondo resonance with the $(\infty,\infty)$ fixed
point. \\
{\bf Summary}: The novel achievements are: i) In composite quantum
dots, such as the TQD displayed in Fig. \ref{TQD}, the two-channel
(left-right leads) Kondo Hamiltonian  (\ref{Hs}) emerges in which
the impurity is a {\it real} spin and the current is due solely to
co-tunneling. The corresponding exchange constant $J_{lr}$ is a
relevant parameter: by taking even and odd combinations, the
system is mapped on an anisotropic two-channel Kondo problem where
$J_{lr}$ determines the degree of anisotropy. ii) RG equations
(\ref{RG3}) for the exchange constants are solved (\ref{RGS}) and
yield the flow diagram displayed in Fig.\ref{flow}. iii)
Although the generic two-channel Kondo fixed-point is not achievable
in the strong coupling limit, inspecting the conductance
$G(V_{ga},T)$ as function of temperature (Fig.\ref{GT}) and
gate voltage (Fig.\ref{GV}) suggests an experimentally
controllable detection of its precursor in the weak 
and intermediate coupling
regimes. iv) Analysis of the Kondo effect in cases of higher spin
degeneracy of the dot ground state is carried out in relation with
dynamical symmetries. The Kondo physics remains that of a fully
screened impurity $n=2S$, and the corresponding Kondo temperatures
are calculated. v) Electron occupation is intimately related to
the nature of the Kondo physics ($n>2S$ for $N=3$ and $n=2S$ for
$N=2,4$). vi) One remark regarding non-linear response is in
order. If a quantum dot in the two channel Kondo regime is subject
to a {\it finite DC bias} there is a peculiar effect of
oscillatory current response \cite{Coleman1}. The TQD in series
discussed here might then be used
to test it experimentally. \\
{\bf Acknowledgment}: We would like to thank A. Schiller Y. Oreg
and A. Shechter for very helpful discussions. Support from BSF,
ISF and DIP grants is highly appreciated.

\end{document}